\documentclass[journal]{IEEEtran}

\usepackage{cite}
\usepackage[pdftex]{graphicx}
\usepackage{amssymb}
\usepackage[cmex10]{amsmath}
\usepackage{amsmath,amsfonts,amssymb}
\usepackage{lineno}
\usepackage{subfigure}
\usepackage{xcolor}
\usepackage{multirow}
\usepackage{diagbox}
\usepackage{threeparttable}
\usepackage{booktabs}
\usepackage{float}
\usepackage{stfloats}
\usepackage{caption}
\usepackage{hyperref}

\begin{document}
	
\title{Grant-Free NOMA-OTFS Paradigm: Enabling Efficient Ubiquitous Access for LEO Satellite Internet-of-Things}

\author{
	Zhen Gao$^*$, Xingyu Zhou$^*$, Jingjing Zhao, Juan Li, Chunli Zhu$^{\dagger}$, Chun Hu$^{\dagger}$, Pei Xiao,
	Symeon Chatzinotas, \\
	Derrick Wing Kwan Ng,~\IEEEmembership{Fellow,~IEEE}, and 
	Bj\"{o}rn Ottersten,~\IEEEmembership{Fellow,~IEEE} 
	
	\thanks{$^*$: The authors contribute equally. }
	\thanks{$^{\dagger}$: Corresponding authors. }
	\thanks{Z. Gao, X. Zhou, J. Li, C. Zhu, and C. Hu are with  with
		the Advanced Research Institute of Multidisciplinary Science, Beijing Institute of Technology, Beijing 100081, China (email:
		\{gaozhen16, zhouxingyu21, jli, chunlizhu, bit\_hc\}@bit.edu.cn).}
	\thanks{J. Zhao is with the Research Institute for Frontier Science, Beihang University, Beijing 100191, China (email: jingjingzhao@buaa.edu.cn).}
	\thanks{P. Xiao is with the Home for 5G and 6G Innovation Centre, Institute for Communication Systems (ICS), University of Surrey, Guildford, GU2 7XH Surrey, U.K. (email: p.xiao@surrey.ac.uk).}
	\thanks{S. Chatzinotas and B. Ottersten are with Interdisciplinary Centre for Security, Reliability and Trust, University of Luxembourg, 29 Avenue J.F. Kennedy, Luxembourg City L-1855, Luxembourg (e-mail: \{symeon.chatzinotas, bjorn.ottersten\}@uni.lu).}	
	\thanks{D. W. K. Ng is with the School of Electrical Engineering and
	Telecommunications, University of New South Wales, Sydney, NSW 2052,
	Australia (e-mail: w.k.ng@unsw.edu.au).}
}


\maketitle

\begin{abstract}
	
With the blooming of Internet-of-Things (IoT), we are witnessing an explosion in the number of IoT terminals, triggering an unprecedented demand for ubiquitous wireless access globally.
In this context, the emerging low-Earth-orbit satellites (LEO-SATs) have been regarded as a promising enabler  
to complement terrestrial wireless networks in providing ubiquitous connectivity and bridging the ever-growing digital divide in the expected next-generation wireless communications.
Nevertheless, the harsh conditions posed by LEO-SATs have imposed significant challenges to the current multiple access (MA) schemes and led to an emerging paradigm shift in system design.
In this article, we first provide a comprehensive overview of the state-of-the-art MA schemes and investigate their limitations in the context of LEO-SATs.
To this end, we propose a novel next generation MA (NGMA), which amalgamates the grant-free non-orthogonal multiple access (GF-NOMA) mechanism and the orthogonal time frequency space (OTFS) waveform, for simplifying the connection procedure with reduced access latency and enhanced Doppler-robustness.
Critical open challenging issues and future directions are finally presented for further technical development.

\end{abstract}

\IEEEpeerreviewmaketitle

\section{Introduction}\label{S1}

\IEEEPARstart{T}{he} Internet-of-Things (IoT) is envisioned to revolutionize our daily life by connecting massive heterogeneous terminals with diverse applications 
and has triggered the evolution of communication technologies in the past decades.  
Meanwhile, the flourishing fifth-generation (5G) communication systems have preliminarily paved the way for the development of IoT vertical domains including smart manufacturing, agriculture, e-health, smart city, and etc \cite{6g-IoT}.
Driven by these demand-intensive vertical industries, the requirement for ubiquitous connectivity with guaranteed quality-of-service (QoS) is more prominent than ever.

\begin{figure*}[t]	
	\centering
	\includegraphics[width=1.5\columnwidth, keepaspectratio]{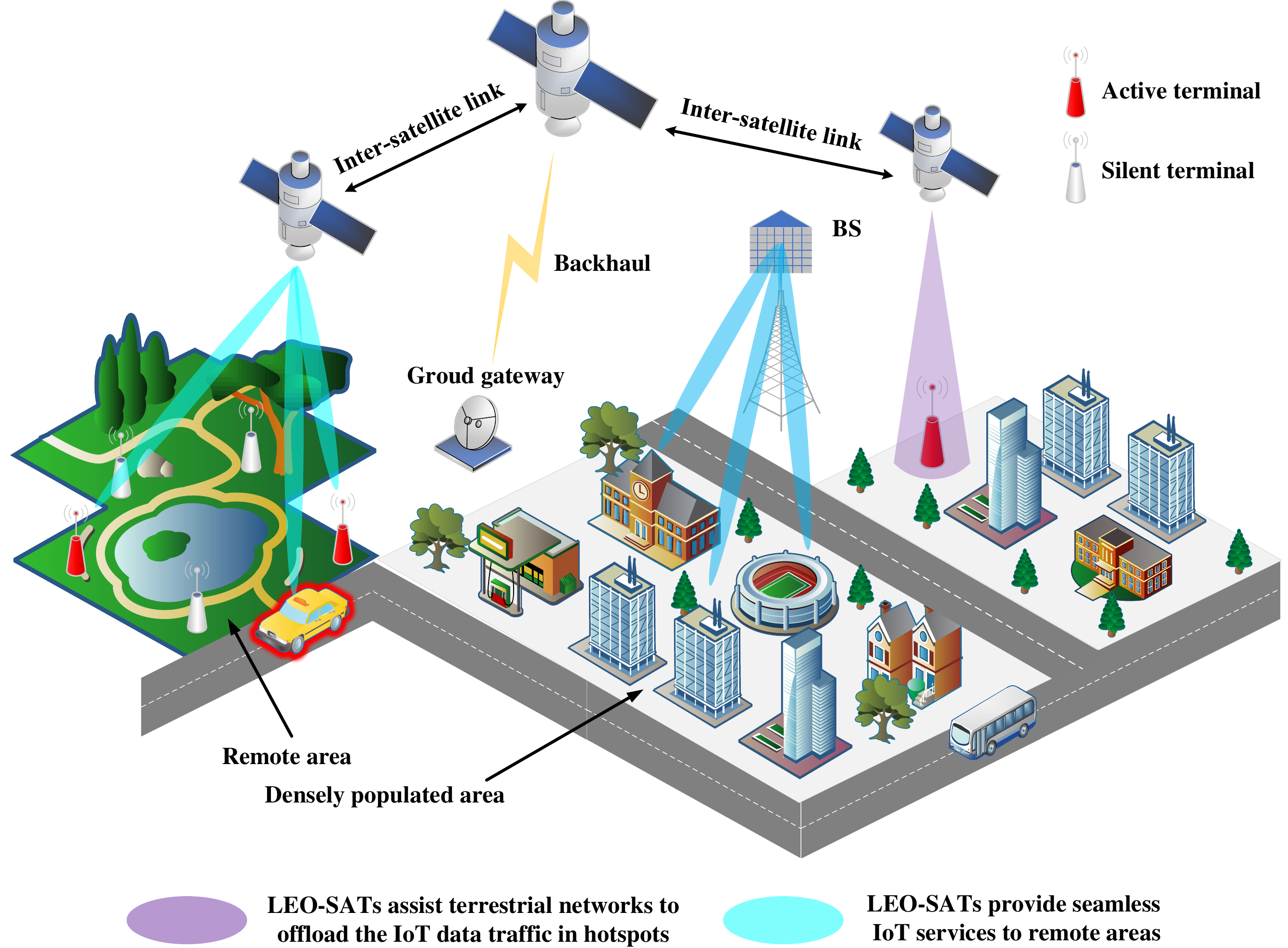}
	\caption{Illustration of the integration of LEO-SATs and terrestrial networks in the expected 6G networks.}	
	\label{fig:scenario}
\end{figure*}

Recently, the escalating data traffic of the IoT has imposed heavy burdens on the existing terrestrial cellular networks. 
As a direct result, the mismatch between the capabilities of 5G networks and the unprecedented IoT demands has been increasingly prominent, especially under the impact of the COVID-19 pandemic \cite{6g-IoT}. 
Moreover, the inherent limitations of terrestrial infrastructures severely restrict their capacity to achieve some anticipated key performance indicators (KPIs), such as access equality, availability, and reliability \cite{SatCon}. 
In this regard, low-Earth-orbit satellites (LEO-SATs) have emerged as a viable solution to supplement and extend terrestrial networks in the expected 6G networks.
By virtue of their global footprints and relatively low round-trip latency, LEO-SATs are regarded as a promising enabler to support ubiquitous global access, bridge the ever-growing digital divide, and cater for increasingly stringent QoS requirements from a multitude of IoT applications, as illustrated in Fig. \ref{fig:scenario}.

Despite the inviting outlook of LEO-SATs, technical innovations are in urgent need to enhance access efficiency at a reasonable cost.
It is predicted that the number of connected IoT terminals grows by 12\%  annually and will eventually reach hundreds of billions with a connection density of 10 million terminals per square kilometer by 2030 \cite{6g-IoT}. 
Against this background,  to fully unleash the underlying potentials of LEO-SATs for accommodating the explosive growth of connections, one of the prerequisites lies in the design of efficient multiple access (MA) paradigms with limited radio resources.
However, it is not straightforward to apply the conventional
MA frameworks to LEO-SATs.
On the one hand, conventional grant-based random access (GB-RA) protocols require complicated handshaking procedures, which would aggravate the access latency, and the induced exceedingly large signaling overheads can impose heavy burdens on the system resources \cite{Grant-free}.
On the other hand, due to the highly dynamic nature of the terrestrial-satellite
link (TSL), traditional MA techniques, which allocate orthogonal or non-orthogonal resource blocks (RBs) in the time-frequency (TF) domain, are likely to suffer from severe  orthogonality impairments due to the imperfect Doppler compensation that jeopardizies the system performance \cite{SatCon}.
To this end, a paradigm shift in MA methodology is indispensable to realize  efficient ubiquitous access for LEO-SATs in practice.

The rest of this article is organized as follows. 
In the following sections, we first provide an overview of the state-of-the-art MA schemes and summarize their limitations for LEO-SATs. Then, we propose a novel next generation MA (NGMA) scheme, which combines grant-free non-orthogonal MA (GF-NOMA) with orthogonal time frequency space (OTFS) modulation to enable efficient ubiquitous access for LEO-SATs.  
Before concluding, important open research issues and future directions are discussed.


\section{Overview of the State-of-the-Art}\label{S2}

Distinguished by different connection procedures, there are mainly two 
categories of MA protocols, namely GB-RA and GF-RA.
These two types of RA protocols have different merits and drawbacks, leading to distinct KPIs including reliability, access latency, and capacity.  
We commence this section by conducting a comprehensive survey of the state-of-the-art MA schemes and summarize them in TABLE \ref{tab:my-table}.
On this basis, we further elaborate their limitations on LEO-SATs and provide guidelines for subsequent research activities.

\renewcommand{\arraystretch}{1.15}
\begin{table*}[]
	\centering
	\caption{A summary of the typical MA schemes in the current literature}
	\label{tab:my-table}
	\begin{threeparttable}
		\resizebox{\textwidth}{!}{%
			\begin{tabular}{|c|c|c|c|c|ccc|l|}
				\hline
				\multirow{2}{*}{\textbf{Solutions}} & \multirow{2}{*}{\textbf{Applicable Channel}}                                                                    & \multirow{2}{*}{\textbf{MA Protocol}}                                                  & \multirow{2}{*}{\textbf{MA Technique}}                       & \multirow{2}{*}{\textbf{Waveform}} & \multicolumn{3}{c|}{\textbf{\begin{tabular}[c]{@{}c@{}}Received Signal \\ Processing\tnote{3} \end{tabular}}} & \multicolumn{1}{c|}{\multirow{2}{*}{\textbf{\begin{tabular}[c]{@{}c@{}}Reasons for incompatibility  \\ with MA in the LEO-SATs\end{tabular}}}}                                  \\ \cline{6-8}
				&                                                                                                                       &                                                                                            &                                                                  &                                             & \multicolumn{1}{c|}{ATI}               & \multicolumn{1}{c|}{CE}                & SD               & \multicolumn{1}{c|}{}                                                                                                                                                                        \\ \hline
				LTE/NR        & \begin{tabular}[c]{@{}c@{}}Frequency selective \\ \& time-varying channels\end{tabular}                               & GB-RA                                                                                & OMA                                                              & (DFT-S)-OFDM\tnote{2}                                & \multicolumn{1}{c|}{}                  & \multicolumn{1}{c|}{\checkmark}        & \checkmark       & \begin{tabular}[c]{@{}l@{}}Complicated access procedures and \\ susceptible to severe Doppler effect\end{tabular}                                                                                   \\ \hline
				\cite{You-LEO}                      & TSLs                                                                                                                  & GB-RA                                                                                & SDMA\tnote{1}                                                             & OFDM                                        & \multicolumn{1}{c|}{}                  & \multicolumn{1}{c|}{}                  & \checkmark       & \begin{tabular}[c]{@{}l@{}}Require the accurate Doppler and \\ delay  compensations at terminals, \\ which is  difficult to implement  in \\ practice\end{tabular}                     \\ \hline
				\cite{Symeon}                       & TSLs                                                                                                                  & GB-RA                                                                                & Frequency-domain OMA                                             & Single-carrier                              & \multicolumn{1}{c|}{}                  & \multicolumn{1}{c|}{\checkmark}        &                  & \begin{tabular}[c]{@{}l@{}}Complicated access procedures \\ aggravate  the access latency and \\ the involved  large  signaling over-\\heads\end{tabular}                             \\ \hline
				\cite{PDMA}                         & \begin{tabular}[c]{@{}c@{}}Broadband time-varying \\ channels\end{tabular}                                            & GB-RA                                                                                & Path-domain OMA                                                  & OTFS                                        & \multicolumn{1}{c|}{}                  & \multicolumn{1}{c|}{\checkmark}        &                  & \begin{tabular}[c]{@{}l@{}}Require excessive orthogonal angle \\ domain resources to assure that the \\ observation  regions for different \\ terminals  do not  overlap\end{tabular}           \\ \hline
				\cite{OTFS-NOMA}                    & \begin{tabular}[c]{@{}c@{}}Broadband time-varying \\ channels\end{tabular}                                            & GB-RA                                                                                & Power-domain NOMA                                                & OTFS                                        & \multicolumn{1}{c|}{}                  & \multicolumn{1}{c|}{}                  & \checkmark       & \begin{tabular}[c]{@{}l@{}}Handshaking procedures and \\ sophisticated  SIC further \\ deteriorate
				 the access  latency\end{tabular} 
				\\ \hline
				\cite{OTFS-SCMA}                      & \begin{tabular}[c]{@{}c@{}}Broadband time-varying\\ channels\end{tabular}                                           & GB-RA                                                                                & Code-domain NOMA                                                 & OTFS                              & \multicolumn{1}{c|}{}        & \multicolumn{1}{c|}{}        & \checkmark       & \begin{tabular}[c]{@{}l@{}}Handshaking procedures and \\ complicated   message passing \\ MUD  algorithm aggravate the \\
				access latency 				
			 \end{tabular}      				                                     \\ \hline
				Globalstar \cite{RA-Sat}            & TSLs                                                                                                                  & SSA                           & Code-domain OMA                                                  & Single-carrier                              & \multicolumn{1}{c|}{}                  & \multicolumn{1}{c|}{}                  & \checkmark       & \begin{tabular}[c]{@{}l@{}}Only accommodate modest traffic \\ and  the packet collisions create \\ potential  network instability issues\end{tabular}                                            \\ \hline
				DVB-RCS \cite{RA-Sat}               & TSLs                                                                                                                  & CRDSA & TDMA\tnote{1}                                                             & Single-carrier                              & \multicolumn{1}{c|}{}                  & \multicolumn{1}{c|}{}                  & \checkmark       & \begin{tabular}[c]{@{}l@{}}Require complicated contention \\ resolution  at the receiver\end{tabular}                                                                                         \\ \hline
				\cite{Yikunmei}                     & \begin{tabular}[c]{@{}c@{}}Braodband quasi-static \\ channels\end{tabular}                                            & GF-RA                                                                                 & Code-domain NOMA                                                 & OFDM                                        & \multicolumn{1}{c|}{\checkmark}        & \multicolumn{1}{c|}{}                  & \checkmark       & \begin{tabular}[c]{@{}l@{}}Suffer from serious performance  \\degradation over the time-varying \\ channels due  to the large interval \\ of the beacon signals and outdated \\ CSI\end{tabular} \\ \hline
				\cite{LEO.IoT}                      & \begin{tabular}[c]{@{}c@{}}Narrowband land mobile \\ satellite (LMS) channels  \\ \end{tabular} & GF-RA                                                                                 & \begin{tabular}[c]{@{}c@{}}Code-domain NOMA\\ +SDMA\end{tabular} & Single-carrier                              & \multicolumn{1}{c|}{\checkmark}        & \multicolumn{1}{c|}{\checkmark}        &                  & \begin{tabular}[c]{@{}l@{}}For narrowband systems only and \\ suffer from low  transmission rate\end{tabular}                                                                                        \\ \hline
				\cite{KML}                          & \begin{tabular}[c]{@{}c@{}}Broadband block-fading \\ channels\end{tabular}                                            & GF-RA                                                                                 & \begin{tabular}[c]{@{}c@{}}Code-domain NOMA\\ +SDMA\end{tabular} & OFDM                                        & \multicolumn{1}{c|}{\checkmark}        & \multicolumn{1}{c|}{\checkmark}        &                  & \begin{tabular}[c]{@{}l@{}}Severe Doppler effect leads to the  \\ orthogonality failure among OFDM \\ sub-carriers\end{tabular}                                                                                                       \\ \hline
				Proposed                            & \begin{tabular}[c]{@{}c@{}}Broadband fast \\ time-varying channels\end{tabular}                                       & GF-RA                                                                                 & \begin{tabular}[c]{@{}c@{}}Code-domain NOMA\\ +SDMA\end{tabular} & OTFS                                        & \multicolumn{1}{c|}{\checkmark}        & \multicolumn{1}{c|}{\checkmark}        & \checkmark       & \multicolumn{1}{c|}{\diagbox{$\quad\quad\quad$}{$\quad\quad\quad\quad\quad\quad$}}                                                                                                                                                                   \\ \hline
			\end{tabular}%
		}
		\begin{tablenotes}    
			\footnotesize               
			\item[1] SDMA: Spatial division MA; TDMA: Time division MA.
			\item[2] DFT-S-OFDM: Discrete Fourier Transform-Spread OFDM.
			\item[3] ATI: Active terminal identification; CE: Channel estimation; SD: Signal detection.			
		\end{tablenotes} 
	\end{threeparttable}
\end{table*}

\subsection{GB-RA Protocols-Based Schemes}

The GB-RA protocols, which were originally conceived for human-type communication, have dominated the existing cellular networks' standards nowadays, e.g., long term evolution (LTE)/LTE-advanced (LTE-A) and 5G new radio (NR).
Besides, they have been adopted by the standardized narrow-band IoT (NB-IoT) to support machine-type communications for co-existing with cellular networks \cite{Grant-free}.
These types of protocols stipulate that the active entities initialize the access procedures through a physical random access channel (PRACH) in a contention-based manner to set up the connection with the base station (BS). 
After the typical four-step handshake procedure is completed, the connected terminals are allowed to initiate a grant acquisition (GA) request for transmitting their payload data to the BS on the allocated orthogonal or non-orthogonal RBs.

In general, these solutions developed for terrestrial cellular networks 
are generally not applicable to LEO-SATs systems due to the harsh channel conditions in the TSLs, such as long propagation delay, severe Doppler effects, strong channel spatial correlation, and etc.
For this purpose, some modifications have been made in recent endeavors \cite{PDMA, OTFS-NOMA, Symeon, You-LEO}. 
For LEO-SATs employing massive multiple-input multiple-output (MIMO) and full frequency reuse techniques, \cite{You-LEO} proposed a space angle-based user grouping algorithm to schedule the served terminals into different groups, where the terminals having different angles of arrival (AoAs) are assigned to the same group and share the same TF resources. While the different groups are allocated to the orthogonal RBs to alleviate the potential interference caused by channel correlation impairment and to maximize the average signal-to-interference-plus-noise ratio.
Without requiring any modification to the current NB-IoT waveform, \cite{Symeon} addressed the issues of MA caused by the typical satellite channel impairments, including long round-trip delays, significant Doppler effects, and wide beams.
Furthermore, \cite{PDMA, OTFS-NOMA,OTFS-SCMA} integrated the OTFS modulation with the GB-RA  protocols and formulated new resource allocation schemes to accommodate the heterogenous mobility demands of differnt IoT applications. 

Despite the aforementioned efforts, some critical issues arise with the sophisticated access procedures in the context of  LEO-SATs-based ubiquitous massive connectivity: 
\begin{itemize}
	\item The access latency (e.g., 14.5 ms \cite{Grant-free}) caused by standard handshakes and grant requests may far exceed the round-trip latency of TSL (e.g., 3 ms \cite{Symeon}) and constitutes as the main latency source.
	
	\item The scheduling signaling overheads are prone to exceed the length of the short-packet payload data generated by IoT applications, leading to excessive waste of resources.
	
	\item With the exponential growth of IoT devices, collisions and network congestion would occur with a high probability due to the limited pools of the orthogonal preamble sequences and the following endless retransmission attempts.	
\end{itemize}
Consequently, these drawbacks of existing GB-RA protocols call for a new paradigm shift to adapt the pivotal features of LEO-SATs and to enable more efficient connections.

\subsection{GF-RA Protocols-Based Schemes} \label{sub}

In this case, the GF-RA protocols arise as a better option to cope with the escalating  IoT traffic demand over satellite communications, where the representative case is ALOHA \cite{SatCon}.
The original ALOHA protocol allows the terminals to transmit the data packets without
any prior coordination, thus only performs well at modest traffic intensity. 
To alleviate the network congestion and the potential network instability issues, more advanced GF-RA techniques have been developed, such as Contention Resolution Diversity ALOHA (CRDSA) and Enhanced Spread Spectrum ALOHA (E-SSA), which integrates diversity or spreading techniques with successive interference cancellation (SIC) \cite{RA-Sat}. 
In fact, Globalstar has deployed the simplex data network based on the SSA-RA protocol
along with the concept of sending multiple replicas of the same burst to
increase reception probability \cite{RA-Sat}.
Meanwhile, some satellite communication standards, such as digital video broadcasting
return channel via satellite (DVB-RCS) and DVB-satellite services to handhelds (DVB-SH), have widely adopted ALOHA-based GF-RA protocols.
Nevertheless, the existing protocols commonly depend on OMA technique and require excessively increasing resources as the number of
terrestrial IoT terminals to be connected grows exponentially.

Recently, an efficient GF-NOMA protocol has been proposed for terrestrial massive machine-type communication (mMTC) \cite{NGMA}. 
This protocol advocates to merge the RA process with data transmission by removing the GA request.   
For instance, when a terminal wakes up, it initializes the pre-configuration and synchronization by exploiting the beacon signals periodically broadcast from the BS. 
Before delivering the uplink payload data, the awakened terminals only need to transmit short non-orthogonal preambles for facilitating the detection of active terminals' set (ATS) at the BS, rather than going through the complicated GA-related procedure.
Furthermore, by integrating the NOMA technique and allocating unique signature patterns in the power, code, and/or spatial domain to different terminals, the GF-NOMA schemes can efficiently separate simultaneously served terminals from the assigned overlapped TF RBs via SIC or multi-user detection (MUD) algorithms \cite{NGMA}. 
At the expense of the complexity at the receiver, the GF-NOMA schemes can significantly reduce the overall latency and signaling overheads with less resource consumption and enhanced access capability.  

The typical GF-NOMA schemes can be divided into two categories based on channel state information (CSI) acquisition: passive \cite{Yikunmei} and active types \cite{KML,LEO.IoT}.
The former passively utilizes the periodically broadcast beacon signals to update the CSI and performs equalization at the terminals or the BS \cite{Yikunmei}. 
This mechanism is effective in quasi-static channels since it saves extra resources to perform channel estimation (CE). 
The latter actively transmits pilot sequences to jointly estimate the ATS and CSI \cite{LEO.IoT}, \cite{KML}. 
As a result, it can better capture the variation of wireless channels, while at the expense of transmission efficiency.

Nevertheless, unlike terrestrial cellular systems, the wireless TSLs have high more dynamic characteristics due to the high mobility of satellite stations and thus robustness to fast time-varying channel conditions is an essential indicator for efficient MA paradigms. In this context, the aforementioned exsiting GF-NOMA schemes still have their limitations:
\begin{itemize}
	\item For the passive-type schemes, Doppler shift can be precompensated at the terminals by exploiting the satellite ephemeris. 
	Yet, it is not a trivial task for the terminals to track the variation of Doppler shift precisely, since beacon signals tend to be broadcasted occasionally to limit the required resource consumption.
	Meanwhile, Doppler precompensation is only precise for terminals located at the beam centre. Inevitably, there will be Doppler residual for others \cite{Symeon}.  
		
	\item For the active schemes, the pilots and data are usually multiplexed in the TF domain, such as orthogonal frequency division multiplexing (OFDM) systems. However, the severe Doppler effect would generally destroy the orthogonality between OFDM subcarriers, which makes it difficult to accurately estimate the ATS and Doppler shift. In this case, traditional techniques could be invalidated as well \cite{KML,LEO.IoT}.
\end{itemize} 
To tackle these pressing issues and motivated by the strengths of burgeoning waveform--OTFS \cite{OTFS-mag}, we intend to incorporate it into the GF-NOMA protocols and heuristically propose a GF-NOMA-OTFS paradigm for LEO-SATs empowering efficient ubiquitous connections in the next section.

\section{Proposed GF-NOMA-OTFS Transmission Paradigm}\label{S3}

OTFS emerges as an alternative to OFDM to accommodate the channel dynamics via modulating information in the delay-Doppler (DD) domain and thus support more reliable services in high-mobility scenarios. 
In this section,  we first review the basic principles of OTFS briefly and discuss its underlying potentials for LEO-SATs. 
Then, we propose to integrate it with the GF-RA mechanism to develop a GF-NOMA-OTFS paradigm and verify its superiority through an illustrative case study.


\subsection{OTFS Waveform}

\subsubsection{Basic Principles}

The essence of OTFS is to shift the data modulation and signal processing from the TF domain to the DD domain.
Applying the two-dimensional (2-D) Fourier transformation to the TF domain channel impulse response (CIR) $h(t,f)$ yields the DD domain CIR $h(\tau,\nu)$.  
By parameterizing the channels with delay $\tau$ and Doppler shift $\nu$, the DD domain CIR can explicitly capture the underlying physics of radio
propagation and directly separate multipath propagations from the delay and Doppler dimensions. 
In sharp contrast to the time-varying $h(t,f)$, the quasi-time-invariant DD domain $h(\tau,\nu)$ enjoys attractive  characteristics, which can be exploited to facilitate efficient signal processing \cite{PDMA}.   
Besides, by attaching a pre-processing and a post-processing modules to the conventional TF domain multi-carrier modulator, OTFS multiplexes data symbols in the DD domain lattices and spreads each one across the whole TF domain via a family of 2-D spreading functions. 
The mathematical structure of OTFS waveform indicates that the impairments caused by double-selective channels, i.e., time- and frequency-shifts, can be represented by 2-D quasi-circular inter-symbol interference (ISI) patterns in the DD domain. 
In fact, the efficient utilization of the DD domain channels and the DD domain multiplexing have endowed OTFS with some unique virtues which render it eminently  suitable for high-mobility scenarios \cite{OTFS-mag}.

\subsubsection{Potentials for LEO-SATs}
   
Incorporating the features of LEO-SATs, we detail the potentials of OTFS as follows.
\begin{itemize}
	\item \textbf{Beneficial features of TSLs in the DD domain:} 
	Although the high mobility of LEO-SATs results in an extremely short coherence time for TSLs in the TF domain, the stability of the DD domain channels is still liable to  maintain. 
	This can be interpreted by the fact that their delay and Doppler parameters, which mainly hinge on the relative locations of satellites and IoT terminals, fluctuate more sluggishly compared with the channel gains in the TF domain.	
	Besides, distinct from the terrestrial propagation environments, there are fewer scatterers in the TSL and the energy of multipath components (MPCs) could be relatively weak owing to the considerable path loss, both of which could substantially contribute to the sparsity of the TSLs in the DD domain.
	Furthermore, as a result of the relatively high altitude of satellite stations compared with the range of the scatterers located in the vicinity of the terminals, all MPCs for the same terminal undergo similar AoA, and thus the DD domain CIR tend to be concentrated along the Doppler dimension.   		
	In short, the TSLs preserve separable, stable, compact, and sparse patterns in the DD domain \cite{OTFS-mag}. 
	
	\item \textbf{Robustness to time-varying channel components:}
	The traditional modulation techniques do not consider the Doppler mitigation, and thus may suffer from severe performance degradation in the time-varying channels. 
	For instance, the popular OFDM modulation efficiently transforms a
	frequency-fading channel to multiple parallel frequency-flat subchannels for low-complexity single-tap equalization.
	However, Doppler shift is treated as a hostile impairment in OFDM, 
	whose orthogonality among different sub-carriers could be destroyed in this case, leading to the inter-carrier interference (ICI).
	By contrast, OTFS is initially conceived for double-selective channels and introduces an additional Doppler dimension to separate the time and frequency fadings in the DD domain, substantially mitigating the destructive effects of fadings' coupling \cite{OTFS-mag}.
	Besides, it is not just the motion of LEO satellite that causes the Doppler shift. 
	The higher frequency oscillators and mismatch between transmitter's and receiver's oscillators inevitably give rise to the issue of undesired phase noise, which introduces significant phase jitters and leads to the equivalently time-varying components.
	In this case, OTFS's robustness against time-varying channels could also be taken as a pivotal enabler to combat the oscillator phase noise, which is critical for high-frequency satellite communications.
	
	\item \textbf{Full channel diversity}:
	The channel diversity benefited from OTFS stems from multiplexing the data symbols in the DD domain and spreading them across the entire TF plane through cascaded OTFS transformations.
	In fact, in contrast to the traditional 1-D modulation scheme such as OFDM,  
	the set of 2-D orthogonal spreading functions of OTFS spans the whole bandwidth and time duration of the transmission packet, which provides additional degrees of freedom in the Doppler (time) dimension. 
	These spreading functions in conjunction with an elaborately designed equalizer in the TF or DD domain allow the exploitation of this full channel diversity gain, which is crucial for performance improvement.
	
	\item \textbf{Other potential benefits}:
	Apart from the aforementioned pivotal features, OTFS also possesses some potential strengths over conventional modulation techniques.   
	For example, OTFS boasts a lower peak-to-average power ratio (PAPR) than OFDM despite its multicarrier nature. Besides, OTFS can significantly reduce pilot and protection overheads, thereby enhancing transmission efficiency, both of which are crucial for resource-limited satellite commnuications.     	 
\end{itemize}



\begin{figure*}[!h]	
	\centering
	\vspace{-5mm}
	\includegraphics[width=1.9\columnwidth, keepaspectratio]{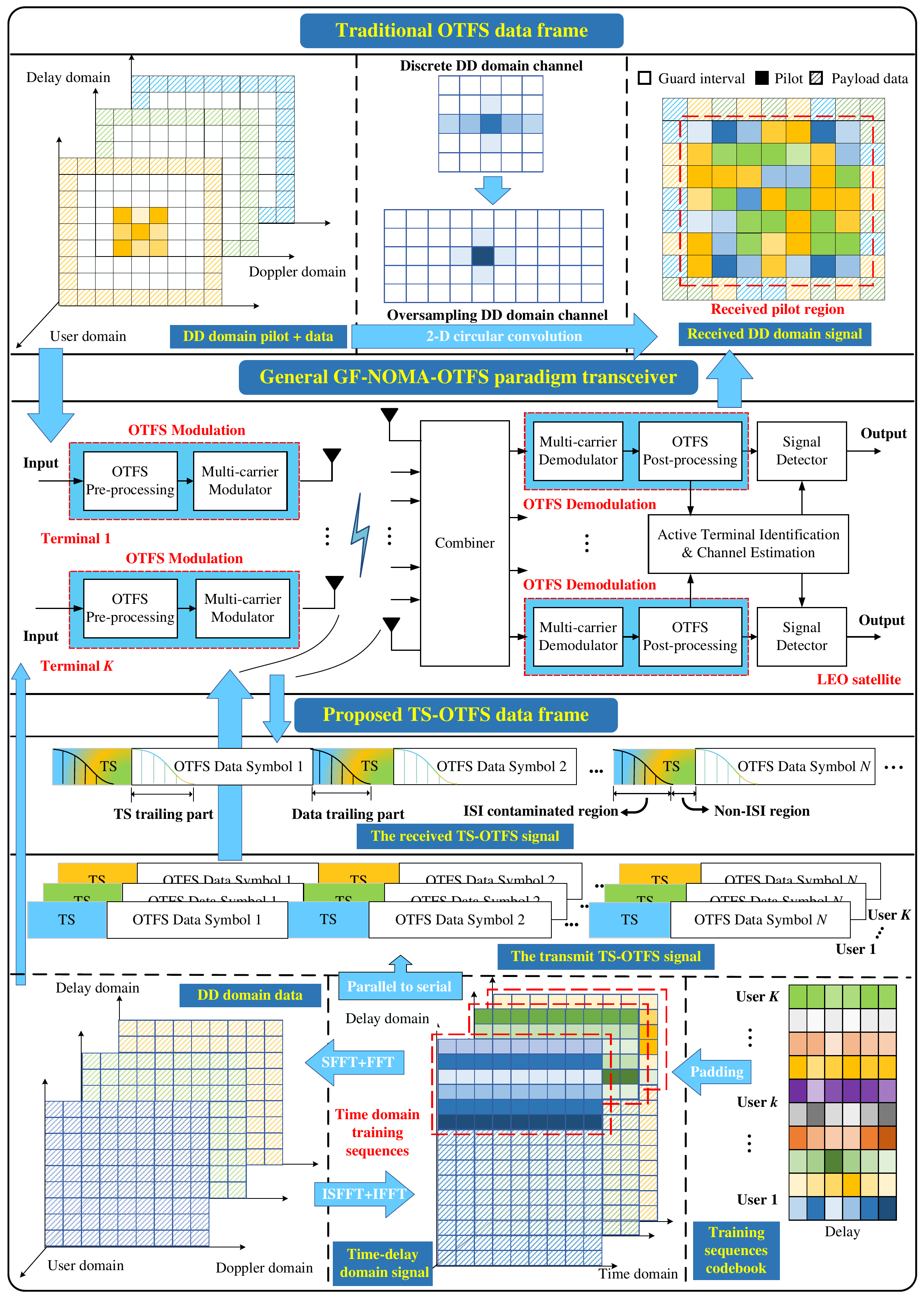}
	\captionsetup{font={color = {black}}, justification = raggedright,labelsep=period}
	\caption{GF-NOMA-OTFS paradigm transceiver and RA signal frame design.}
	\label{fig:sysyem}	
\end{figure*}

\subsection{GF-NOMA-OTFS Paradigm}

As an amalgamation of the GF-RA mechanism with the OTFS waveform, the transmission procedure of the proposed GF-NOMA-OTFS paradigm can be summarized as follows: 
\textit{Step 1)} After initializing the pre-configuration and the time synchronization, the active IoT terminals are allowed to directly transmit
RA uplink signals without complicated access and resource scheduling request; 
\textit{Step 2)} The satellite stations collect signals received from all active terminals and perform active terminal identification (ATI) and CE,
where the estimated ATS and CSI are used for the following DD domain data demodulation in the absence of scheduling information; 
{\textit{Step 3)} The satellite stations feedback the demodulated data through inter-satellite links (ISLs) and backhaul to the gateway and broadcast acknowledgment
character to the active terminals. }
In order to achieve more efficient ATI and CE, we first propose an improved RA signal frame structure for OTFS waveform.

\subsubsection{RA Signal Frame Structure}

The widely adopted OTFS frame structure in the existing literature consists of the DD domain data symbols, guard intervals, and pilot symbols \cite{OTFS-mag} as illustrated in Fig.~\ref{fig:sysyem}.
{Since the pilots are directly placed in the sparsity domain (DD), the direct CE via
the least-square (LS) or maximum likelihood (ML) approach are feasible and the most efficient schemes. 
However, its performance is severely restricted by the OTFS resolution (especially Doppler domain) \cite{NGMA}.
To further enable compressive sensing (CS)-based and super-resolution CE approach, we propose a training sequence (TS)-aided OTFS (TS-OTFS) frame structure as shown in Fig.~\ref{fig:sysyem}.}
Specifically, a TS-OTFS frame consists of two segments, corresponding to TSs and OTFS payload data, respectively. 
As for the OTFS payload data, the raw input bit streams generated by IoT terminals are firstly multiplexed in the DD domain, with delay and Doppler dimensions being denoted as $M$ and $N$, respectively.  
Then, the DD domain data is transformed into the time domain signal through a cascade of an OTFS pre-processing module and a multi-carrier modulator. 
{As for the TSs, they are pre-allocated non-orthogonal random (e.g., pseudo-noise or complex Gaussian) sequences with low-correlation placed in the time-delay domain as unique identifiers for different terminals, whose length $M_t$ is far smaller than the total number of served terminals (OMA cases).
$N$ identical TSs are embedded into OTFS payload data converted to the time domain as the substitute for the DD domain pilots to facilitate ATI and CE.}
Next, we leverage the TSs and the characteristics of TSL in the DD domain mentioned above to implement ATI and CE.

\subsubsection{ATI and CE Algorithm}

In the presence of propagation delay of MPCs, TSs are doomed to undergo severe ISI from the unknown payload data preceding it. 
In order to overcome the ISI impact on the ATI and CE, an effective approach is to utilize the non-ISI region of TSs as illustrated in Fig. \ref{fig:sysyem}, which is the tail-end of TSs and thereby merely depends on the known transmit TSs. 

To begin with, by exploiting the 1-D convolution relationship between the received TSs and the transmit TSs in the delay domain, and the superposition of TSs from different active terminals, 
the first stage of ATI and CE can be formulated as a typical sparse signal recovery problem, where the few non-zero elements of delay domain channels of active terminals are required to be recovered.
In this way, by means of cutting-edge CS algorithms, the delay domain CIR vectors of all potential terminals can be recovered with the low-dimensional non-ISI region so that the length of TSs overhead can be substantially reduced.
In line with the estimated channel gain, a power threshold detecor could be further utilized to decide activity for each potential terminal, and thus acquire ATS \cite{KML} . 

As for CE, the accurate reconstruction of the DD domain CIR  $h(\tau,\nu)$ lies on the acquisition of key channel parameters, i.e., delay, Doppler shift, and fading coefficients. 
Specifically, the support set of recovered delay domain sparse CIR vectors can straightforwardly map to the delay parameters. 
Furthermore, since the number of Doppler component $L$ tends to meet $L \ll N$, 
by collecting the recovered delay domain sparse CIR vectors corresponding to the $N$ TSs ($N$ different instants), the Doppler parameters can be estimated with super-precision in the time-domain resorting to the spatial spectrum estimation algorithm, e.g., estimation of signal parameters using rotational invariance techniques (ESPRIT), and 
multiple signal classification (MUSIC).
Eventually, the fading coefficients associated with each physical propagation path can be mathematically determined based on the LS criterion.  

\color{black}

\begin{table}[]
	\centering
	\captionsetup{font={color = {black}}, justification = raggedright,labelsep=period}
	\caption{Simulation parameter settings}	
	\label{tab:my-table2}
	\resizebox{0.85\columnwidth}{!}{%
		\begin{tabular}{|c|c|}
			\hline
			\textbf{Parameter}                      & \textbf{Value}   \\ \hline
			Carrier frequency {[}GHz{]}             & 10               \\ \hline
			System bandwidth {[}MHz{]}              & 122.88           \\ \hline
			Subcarrier spacing {[}kHz{]}            & 480              \\ \hline
			OTFS grid (delay $\times$ Doppler)      & $256 \times 8$   \\ \hline
			Satellite antenna array size  & \begin{tabular}[c]{@{}c@{}}$32 \times 32$ UPA \\ digital beamforming\end{tabular} \\ \hline
			Terminal antenna array size   & \begin{tabular}[c]{@{}c@{}}$32 \times 32$ UPA \\ analog beamforming\end{tabular}  \\ \hline
			Quantizer (precision: $B^{\rm adc}$)    & Lloyd-Max        \\ \hline
			Number of potential terminals          & 1000             \\ \hline
			Activation probability                  & 0.01             \\ \hline
			Satellite orbital altitude {[}km{]}     & 500              \\ \hline
			Zenith angle range $[^\circ]$           & $[-44.7,44.7]$   \\ \hline
			Azimuth angle range $[^\circ]$          & $[0,360)$        \\ \hline
			Doppler shift range {[}kHz{]}           & $[-178.2,178.2]$ \\ \hline
			Atmospheric loss {[}dB{]}               & 0.07             \\ \hline
			Scintillation loss {[}dB{]}             & 2.2              \\ \hline
			Shadowing margin {[}dB{]}               & 3                \\ \hline
			Free space path loss {[}dB{]} & $52.45+20\log_{10}(d[m])$                                                         \\ \hline
			Received signal-to-noise ratio (SNR) {[}dB{]} & 15               \\ \hline
			Number of MPCs for each TSL             & 2                \\ \hline
		\end{tabular}%
	}
\vspace{-3mm}
\end{table}

\begin{figure*}[t]	
	\centering
	{\subfigure[]{
			\includegraphics[width=0.65\columnwidth, keepaspectratio]{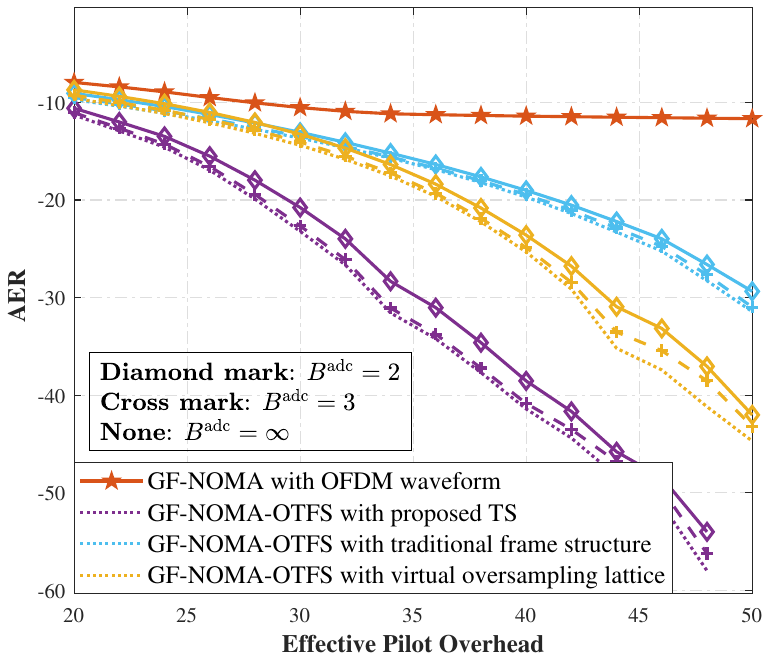}
			\label{fig:AER}
		}
		\subfigure[]{
			\includegraphics[width=0.645\columnwidth, keepaspectratio]{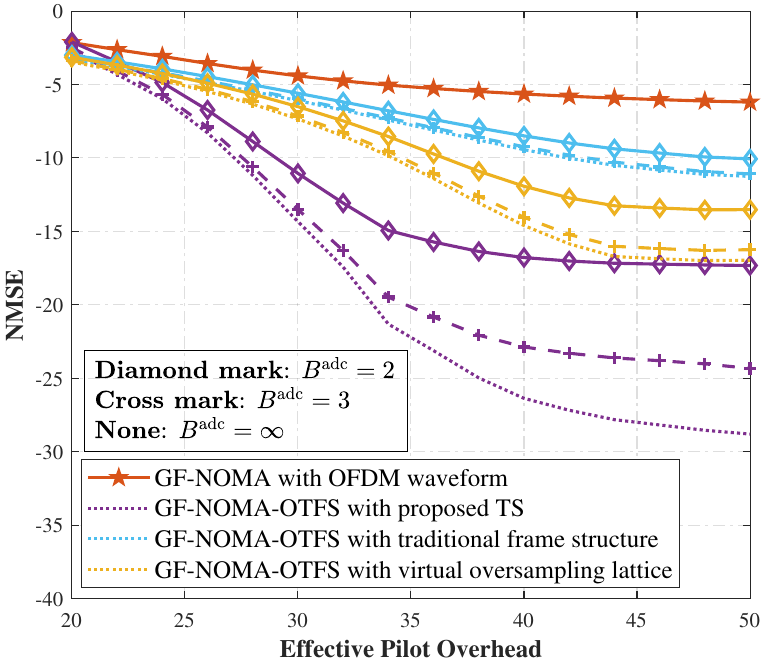}
			\label{fig:NMSE}
		}
		\subfigure[]{
			\includegraphics[width=0.64\columnwidth, keepaspectratio]{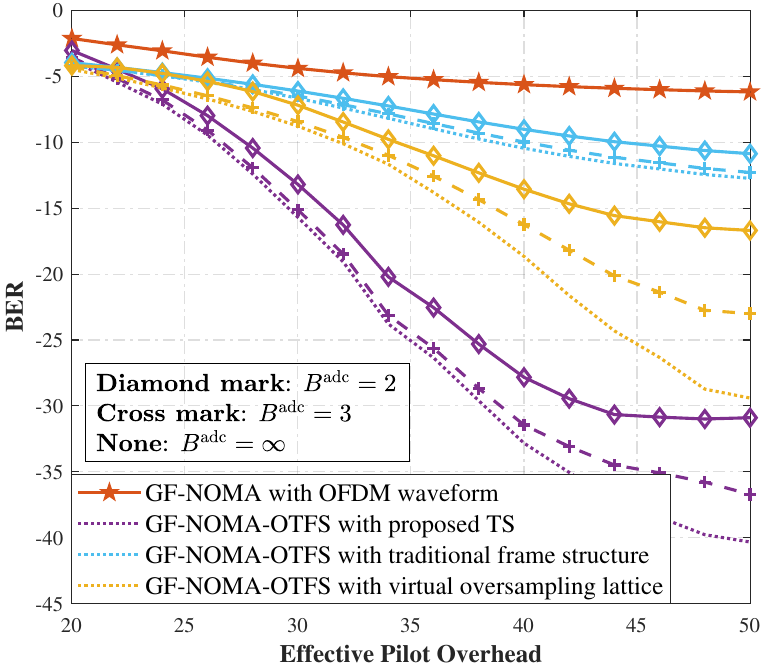}
			\label{fig:BER}
	}}
	\caption{Performance comparison of different MA schemes:
		(a) AER of ATI; (b) NMSE of CE; (c) BER of SD.}
	\label{fig:simulation}
\end{figure*}

\subsection{Case Study}

{To compare the existing MA schemes surveyed in Section II and  
demonstrate the superiority of the GF-NOMA-OTFS paradigm and the frame structure design, we study a representative case in the LEO-SATs-based massive IoT access and provide simulation results.
For comparison, we consider the following schemes:
\begin{itemize}
	\item
	\textbf{GF-NOMA with OFDM waveform \cite{KML}}: the uplink RA signals utilize frequency-domain pilots and space-frequency structured sparsity of access channel matrices to perform ATI and CE. Its effective pilot length is defined as the time slots occupied by frequency-domain pilots.     
	
	\item  
	\textbf{GF-NOMA-OTFS with traditional frame structure \cite{OTFS-mag}}: 
	the traditional OTFS RA signals utilize DD domain embedded pilots and its 2-D circular convolution interaction with discrete DD structured-sparse access channel matrices to perform ATI and CE, cf. Fig. \ref{fig:sysyem}. Its effective pilot length is defined as the size of pilot block along delay dimension. 
	
	\item  
	\textbf{GF-NOMA-OTFS with virtual oversampling lattice \cite{OTFS-mag}}: 
	the virtual oversampling representation is further utilized to improve virtual Doppler and delay resolutions and mitigate access channel spreading due to discretization, cf. Fig. \ref{fig:sysyem}. 
	
	\item
	\textbf{GF-NOMA-OTFS with proposed TS}:
	the uplink RA signals utilize proposed TS and DD domain sparsity of access channel matrices to perform ATI and CE. Its effective pilot length is defined as the non-ISI region of TS.		
\end{itemize}

The simulation parameter settings are summarized in TABLE \ref{tab:my-table2}. 
It is noteworthy that the LEO satellite is expected to be equipped with a large phased array to enhance beamforming capability, which requires numerous radio frequency components including highly linear power amplifier and analog-to-digital converter (ADC).
Therefore, we consider to employ low-resolution ADCs for receiver of LEO satellite, which enjoy low power consumption and simple hardware implementation, to meet the on-board power and cost constraints.


Fig. \ref{fig:AER} and Fig. \ref{fig:NMSE} respectively depict the activity error rate (AER) and normalized mean square error (NMSE) of different schemes as a function of effective pilot length.
Specifically, the severe Doppler effect incurs ICI for the GF-NOMA scheme employing OFDM waveform and thus dominates significant performance loss regardless of pilot overhead.
In contrast, the proposed GF-NOMA-OTFS paradigms enjoy performance superiority by handling the Doppler effect with their DD domain signal processing manner and the exploitation of the DD domain channel's property.
Unfortunately, due to the impact of discretization, the channel spreading and the limited resolution destroy the beneficial features of TSLs in the DD domain, degrading its optimal performance. 
The utilization of virtual oversampling lattice can compensate the performance loss to some extent while at the cost of higher computation complexity. 
Moreover, the proposed TSs enable the combination of CS-based estimation in the delay domain and super-resolution estimation in the time domain, ensuring superior performance while keeping lower pilot overheads.   

On this basis, we further compare the performance of bit error rate (BER) in Fig. \ref{fig:BER}
based on the estimated ATS and CSI, where a straightforward DD domain LS multi-user signal detector is utilized to perform SD.  
The performance gain achieved by the proposed GF-NOMA-OTFS paradigm suggests the noticeable benefits from more accurate ATS and CSI, and the full channel diversity extraction provided by OTFS modulation.
Furthermore, it is noteworthy that with the limited resolution ADC, i.e., $B^{\rm adc} = 2,3 \, \rm bits$, the performance of the GF-NOMA-OTFS paradigms degrades sluggishly in contrast to ideal ADC, which indicates the proposed schemes remain robust in this case, making  hardware-efficient implementation possible for satellite payloads.

\color{black}
\section{Open Issues and Future Directions}\label{S4}

As a fledgling concept, the GF-NOMA-OTFS paradigm not only unveils opportunities but also poses challenges. 
In this section, we discuss the challenges to be addressed and highlight the future research directions to spur more technological breakthroughs in the future.

\subsection{Terrestrial and On-Board Constraints Optimization}

The IoT terminals are expected to be power-constrained for low carbon footprint and cost-efficient deployment, which yet may result in failure to close the link considering the significant path loss.
At the cost of achievable data rate, the repetition code spreading specified in NB-IoT standard \cite{Symeon}, and the direct sequence spread spectrum widely used in satellite communications \cite{OTFS-DS}, can be integrated with the OTFS modulation to improve link SNR and enhance robustness with low transmit power. 
In addition, imperfect hardware components remain another challenge. \color{black}
Practical satellite communication systems usually work within the saturation point of power amplifiers to provide the most efficient power output and compensate the significant path loss, yet results in severe non-linear behaviors.
In particular, most multi-carrier waveforms are with inconstant signal envelopes, which is likely to lead to non-linear distortion in this case.
On the other hand, low-resolution digital-to-analog converters and ADCs, have attracted increasing research attention to be equipped for terminal and satellite apparatus.
In the presence of non-ideal characteristics, undesired interference is often inevitable.
It is suggested that the joint optimization of performance taking these on-board and terrestrial constraints into account is an underlying trend in the future research of MA paradigms \cite{KML}.

\subsection{Interplay with Other NGMA Schemes}

In addition to SDMA, other cutting-edge NOMA paradigms, e.g., power-domain NOMA \cite{OTFS-NOMA} and sparse code MA \cite{OTFS-SCMA}, have been considered to integrate with OTFS waveform to enhance spectral efficiency and alleviate the spatial correlations caused by overloaded SDMA. 
Therefore, it is rewarding to investigate their amalgamation with the proposed MA scheme for more efficient payload data multiplexing, striking a tradeoff between the spectral efficiency and transmission reliablity.
In this case, SIC and MUD algorithms are pivotal but challenging in the presence of 
more complicated multi-dimensional (Doppler, delay, spatial domains) interweaving and inter-users interference. 
In fact, under the framework of single-input single-output OTFS, a myriad of SD schemes have been proposed from different perspectives \cite{OTFS-mag}, including but not limited to execution in various domains, e.g., the DD domain, the TF domain, and the cross-domain iteration; different optimization objective, e.g., lower complexity and superior BER performance; and different techniques, e.g.,  zero forcing, minimum mean square error, ML, and message passing.
On this basis, their extension to more sophisticated multi-user NOMA systems should be further studied. 

\color{black}
\subsection{NGMA for Cooperative Networks }

Benefiting from the ultra-dense network topology, the concept of multi-connectivity arises in LEO-SATs networks \cite{SatCon}. 
{In this case, each terminal can be covered by multiple  satellite stations simultaneously, and thus more available connection options arise:
an IoT terminal may communicate with the preferred satellite for the shortest link 
or connect to multiple satellites concurrently with the aid of ISLs for exploiting diversity gain and avoiding frequent handover. }
Besides LEO-SATs, assortments of non-terrestrial infrastructures, such as high altitude platforms and unmanned aerial vehicles are expected to be integrated with the terrestrial interfaces to constitute an integrated space-air-ground-sea network for 6G communications \cite{KML}. 
Therefore, how to coordinate the intra-network and inter-network cooperation with the NGMA scheme to facilitate the seamless integration of such  multi-dimensional heterogeneous networks should be further investigated for QoS enhancement. 
To this end, key technologies, such as interference management, cognitive radio, and software-defined networking \cite{SatCon}, which show promise to cope with more complicated interference scenarios among different interfaces and contribute to more flexible network architectures, are worthy of further exploration. 

%

   
\color{black}   
              
\section{Conclusions}\label{S7}

Efficient NGMA paradigms play a critical role for LEO-SATs in the expected 6G networks to fully unleash their potentials and support ubiquitous massive connectivity.
We commenced this article by providing a comprehensive overview of the state-of-the-art MA schemes and highlighting their limitations in the context of LEO-SATs.
By amalgamating the advantages of the GF-RA mechanism and the OTFS waveform, we proposed a GF-NOMA-OTFS paradigm to provide more efficient access and accommodate the high-mobility of TSLs.
The case study validated that the proposed solution offered prominent gains for ATI and CE in the absence of scheduling information, while harvesting BER performance improvement by fully exploiting full channel diversity.    
Yet, there remain a plethora of open challenges to be addressed, 
encouraging more technological breakthroughs in the future.

\end{document}